\documentclass[11pt,a4paper]{article}
\usepackage[hyperref]{acl2020}
\usepackage{times}
\usepackage{latexsym}

\usepackage{graphicx}
\usepackage{tabularx}
\usepackage{soul}

\usepackage{epstopdf}
\usepackage[latin1]{inputenc}

\usepackage{hyperref}
\usepackage{xstring}

\usepackage{xcolor}
\usepackage{epstopdf}

\usepackage{microtype}

\aclfinalcopy 

\title{Examining Citations of Natural Language Processing Literature}

\author{Saif M. Mohammad\\
National Research Council Canada\\
Ottawa, Canada\\ 
\texttt{saif.mohammad@nrc-cnrc.gc.ca.}\\}

\date{}

\begin{document}
\maketitle
\begin{abstract}
We extracted information from the ACL Anthology (AA) and Google Scholar (GS) to examine trends in citations of NLP papers.
We explore questions such as: how well cited are papers of different types (journal articles, conference papers, demo papers, etc.)?
how well cited are papers from different areas of within NLP? etc.
Notably, we show that only about 56\% of the papers in AA are cited ten or more times.
CL Journal has the most cited papers, but its citation dominance has lessened in recent years.
On average, long papers get almost three times as many citations as short papers;
and papers on \textit{sentiment classification},  \textit{anaphora resolution}, and \textit{entity recognition} have the highest median citations.
The analyses presented here, and the associated dataset of NLP papers mapped to citations, have a number of uses including:  
understanding how the field is growing and 
quantifying the impact of different types of papers. 
\end{abstract}

\section{Introduction}
The origins of Natural Language Processing (NLP) go back to the earliest work in Computer Science---when Alan Turing
published his seminal paper exploring whether machines can think, and proposed what is now known as the \textit{Turing test} \cite{machinery1950computing,turing2009computing}.
A crucial factor in the evolution of NLP as a field of study in its own right was the formation of the Association for Computational Linguistics (ACL) in 1962, and the first ACL conference in 1965.\footnote{One can make a distinction between NLP and Computational Linguistics; however, for this work, we will consider them to be synonymous. Also, ACL was originally named the Association for Machine Translation and Computational Linguistics (AMTCL). It was changed to ACL in 1968.}  
Today NLP is a broad interdisciplinary field with a growing number of researchers from Computer Science, Linguistics, Information Science, Psychology, Social Sciences, Humanities, and more joining its ranks.

Organizations such as ACL, ELRA, and AFNLP publish peer-reviewed NLP papers that include both journal articles and 
conference proceedings. Historically, the need for a faster review process has made conference proceedings the
dominant form of published research in Computer Science and NLP.
With time, the conferences and the types of papers they publish, have  evolved.
Some conferences, such as  EMNLP and ACL, are highly competitive, while others, such as most workshops and LREC,
deliberately choose to keep more generous acceptance rates. The publications themselves can be of different types:
journal articles, conference papers, short papers, system demonstration papers, shared task papers,
workshop papers, etc.
New ideas and paradigms have evolved: for example, the rise of statistical NLP in the 1990s and deep learning in the 2010s.
With the dawn of a new decade and NLP research becoming more diverse and more popular than it ever has been,
this work looks back at the papers already published to identify broad trends 
in their impact on subsequent scholarly work.  


Commonly used metrics of research impact on subsequent scholarly work are derived from citations including: number of citations, average citations, h-index, relative citation ratio, and impact factor \cite{bornmann2009state}.
However, the number of citations is not always a reflection of the quality or importance of a piece of work. Note also that there are systematic biases that prevent certain kinds of papers from accruing citations, especially when the contributions of a piece of work are atypical or in an area where the number of scientific publications is low. Furthermore, the citation process can be abused, for example, by egregious self-citations \cite{ioannidis2019standardized}.
Nonetheless, given the immense volume of scientific literature, the relative ease with which one can track citations using services such as Google Scholar (GS), 
and given the lack of other easily applicable and effective metrics, citation analysis is an imperfect but useful window into research impact.

Thus citation metrics are often a factor when making decisions about funding research and hiring scientists. Citation analysis can also be used to gauge the influence of outside fields on one's field and the influence of one's field on other fields. 
Therefore, it can be used to determine the relationship of a field with the wider academic community.

As part of a broader project on analyzing NLP Literature,
we extracted and aligned information from the ACL Anthology (AA) and Google Scholar to create a dataset of tens of thousands of NLP papers and their citations \cite{mohammad2020data,mohammad2019nlpscholar}.\footnote{In separate work we have used the NLP Scholar data to explore gender gaps in Natural Language Processing research;
especially, disparities in authorship and citations \cite{mohammad2020gender}.
We have also developed an interactive visualization tool that allows users to search for relevant related work in the ACL Anthology \citet{mohammad2020demo}.}
In this paper, we describe work on examining the 
papers and their citations to identify broad trends within NLP research---overall, across paper types, across publication venues, over time, and across research areas within NLP. 
Notably, we explored questions such as: how well cited are papers of different types (journal articles, conference papers, demo papers, etc.)?
how well cited are papers published in different time spans?
how well cited are papers from different areas of research within NLP? etc.
The dataset and the analyses have many uses including:  
understanding how the field is growing; 
quantifying the impact of different types of papers on subsequent publications;
and understanding the impact of various conferences and journals.
Perhaps most importantly, though, 
 they serve as a record of the state of NLP literature in terms of citations.
All of the data and interactive visualizations associated with this work are freely available through the project homepage.\footnote{http://saifmohammad.com/WebPages/nlpscholar.html} 

\section{Background and Related Work}

The ACL Anthology is a digital repository of public domain, free to access, articles on NLP.\footnote{https://www.aclweb.org/anthology/} It includes papers published in the family of ACL conferences as well as in other NLP conferences such as LREC and RANLP.\footnote{ACL licenses its papers with a Creative Commons Attribution 4.0 International License.} 
As of June 2019, it provided access to the full text and metadata for $\sim$50K articles published since 1965 (the year of the first ACL conference).
It is the largest single source of scientific literature on NLP.
Various subsets of AA have been used in the past for a number of tasks including:
the study of citation patterns and intent \cite{pham2003new,aya2005citation,teufel2006automatic,mohammad2009using,nanba2011classification,zhu2015measuring,radev2016bibliometric},
generating summaries of scientific articles \cite{qazvinian2013generating},
and creating corpora of scientific articles \cite{bird2008acl,mariani2018nlp4nlp}.
Perhaps the work closest to ours is that by \newcite{anderson2012towards}, who examine papers from 1980 to 2008 to track 
  the ebb and flow of topics within NLP,
  the influence of subfields on each other,
  and the influence of researchers from outside NLP. 
However, that work did not examine trends in the citations of NLP papers.
  
Google Scholar is a free web search engine for academic literature.\footnote{https://scholar.google.com}
Through it, users can access the metadata associated with an article 
such as the number of citations it has received.
Google Scholar does not provide information on how many articles are included in its database. However, scientometric researchers estimated that it included about 389 million documents in January 2018 \cite{gusenbauer2019google}---making it the world's largest source of academic information. 
Thus, there is growing interest in the use of Google Scholar information to draw inferences about scholarly research  in general \cite{howland2010scholarly,orduna2014size,khabsa2014number,mingers2015review,martin2018google} 
and on scholarly impact in particular \cite{priem2010scientometrics,yogatama2011predicting,bulaitis2017measuring,ravenscroft2017measuring,bos2019interdisciplinary,ioannidis2019standardized}.
This work examines patterns of citations of tens of thousands of NLP papers, both overall and across paper types, venues, and areas of research.
 
\section{Data}
 We now briefly describe how we extracted information from the ACL Anthology  and Google Scholar to facilitate
the citation analysis. (Further details about the dataset, as well as an analysis of the volume of  research in NLP over the years,
are available in \citet{mohammad2020data}.)
We aligned the information across AA and GS using the paper title, year of publication, and first author last name.

 \begin{figure*}[t!]
 \begin{center}
 	\includegraphics[width=2.08\columnwidth]{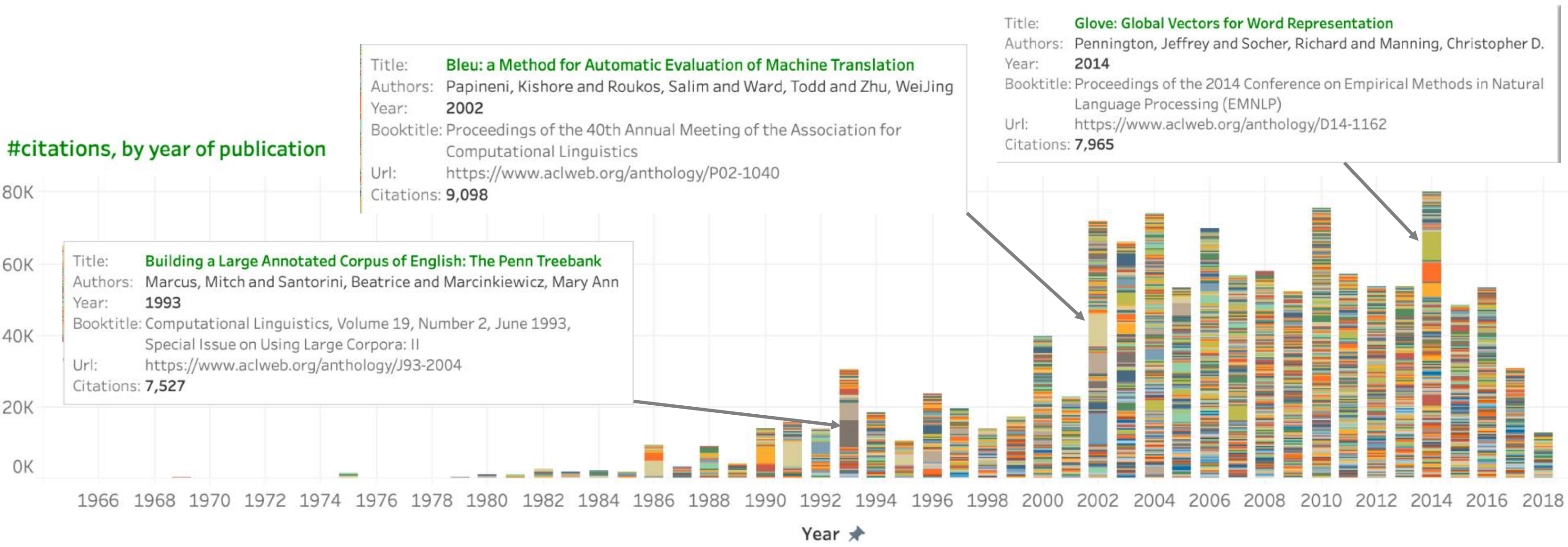}
 	\caption{A timeline graph 
 	of citations received by papers published in each year. Colored segments correspond to papers; the height of a segment is proportional to the number of citations. Hovering over a paper shows metadata.}
 	\label{fig:pubs-timeline-labels}
 \end{center}
 \end{figure*}

\subsection{ACL Anthology Data}
The ACL Anthology provides access to its data through its website and a github repository \cite{gildea-etal-2018-acl}.\footnote{https://www.aclweb.org/anthology/\\https://github.com/acl-org/acl-anthology}
We extracted paper title, names of authors, year of publication, and venue of publication from the repository.\footnote{Multiple authors can have the same name and the same authors may use multiple variants of their names in papers. 
The AA volunteer team handles such ambiguities using both semi-automatic and manual approaches (fixing some instances on a case-by-case basis). 
Additionally, the AA repository includes a file that has canonical forms of author names.}


As of June 2019, AA had $\sim$50K entries; however, this includes forewords, schedules, etc.\@
that are not truly research publications. 
After discarding them we are left with a set of 44,894 papers.\footnote{We used simple keyword searches for terms such as
 {\it foreword, invited talk, program, appendix} and {\it session} in the title to pull out
 entries that were likely to not be research publications. These were then manually examined to verify
 that they did not contain any false positives.}


\subsection{Google Scholar Data}


Google Scholar does not provide an API to extract information about the papers. This is likely because of its agreement with publishing companies that have scientific literature behind paywalls \cite{martin2018google}.  We extracted citation information from Google Scholar profiles of authors who published at least three papers in the ACL Anthology. A Google Scholar Profile page is a user-created page where authors can include their papers (along with the GS-provided citation information for the papers). Scraping author profile pages is explicitly allowed by GS's robots exclusion standard. This is also how past work has studied Google Scholar \cite{khabsa2014number,orduna2014size,martin2018google}.

We collected citation information for 1.1 million papers in total. We will refer to this dataset as 
\textit{GScholar-NLP}. 
Note that GScholar-NLP includes citation counts not just for NLP papers, but also for non-NLP papers published by the authors. 
GScholar-NLP includes 
32,985 of the 44,894 papers in AA (about 74\%). We will refer to this subset of the ACL Anthology papers as AA$'$. The citation analyses presented in this paper are on AA$'$.
Future work will analyze both AA$'$ and GScholar-NLP to determine influences of other fields on NLP.

\section{Examining Citations of NLP Papers}

We use data extracted from the ACL Anthology and Google Scholar to examine trends in citations through a series of questions.\\[-3pt]

\noindent \textit{Q1. How many citations have the AA$'$ papers received? How is that distributed among the papers published in various years?}\\[-5pt]

\noindent A. $\sim$1.2 million citations (as of June 2019). Figure \ref{fig:pubs-timeline-labels} shows the screenshot of an interactive timeline graph where each year has a bar with height corresponding to the number of citations received by papers published in that year. Further, the bar has colored segments corresponding to each of the papers; the height of a segment is proportional to the number of citations the paper has received. Thus it is easy to spot the papers that received a large number of citations. 
Hovering over individual papers reveals additional metadata.\\[-9pt]

\noindent \textit{Discussion:} With time, not only have the number of papers grown, but also the number of high-citation papers. We see a marked jump in the 1990s over the previous decades, but the 2000s are the most notable in terms of the high number of citations. The 2010s papers will likely surpass the 2000s papers in the years to come.\\[-3pt]

\noindent \textit{Q2. How well cited are individual AA$'$ papers, as in, what is the average number of citations, what is the median, what is the distributison of citations?
How well cited are the different types of papers: journal papers, main conference papers, workshop papers, 
etc.?}\\[-5pt]

\begin{figure}[t!]
 \begin{center}
 	\includegraphics[width=\columnwidth]{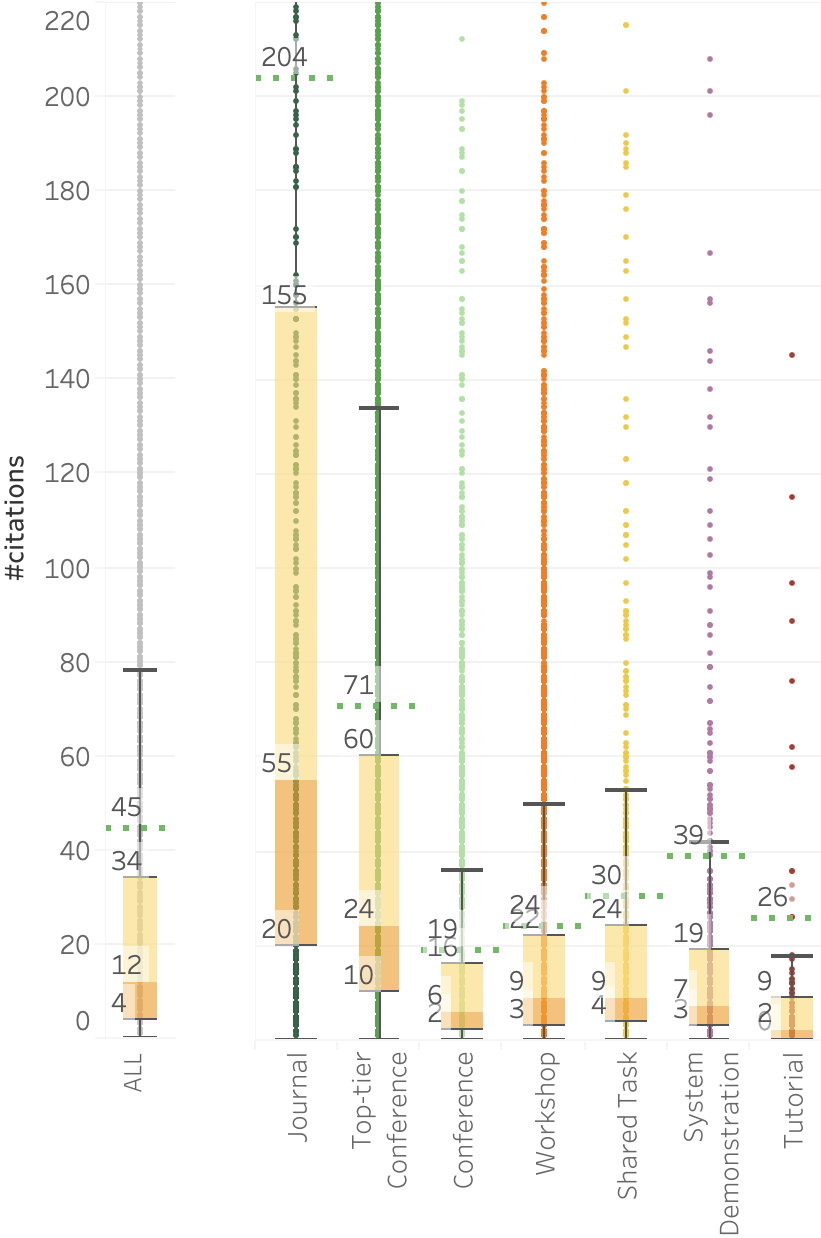}
 	\caption{Citation box plots for papers published 1965--2016: overall and by type.}
 	\label{fig:citn-overall-ptype}
 \end{center}
 \end{figure}

\noindent A. In this and all further analyses, we do not include AA$'$ papers published in 2017 or later (to allow for at least 2.5 years for the papers to collect citations). There are 26,949 AA$'$ papers that were published from 1965 to 2016.
Figure \ref{fig:citn-overall-ptype} shows box and whisker plots for:
all of these papers (on the left) and for individual paper types (on the right).
The whiskers are at a distance of 1.5 times the inter-quartile length. 
The average number of citations are indicated with the horizontal green dotted lines.
Creating a separate class for ``Top-tier Conference" is somewhat arbitrary, but it helps make certain comparisons more meaningful. For this work, we consider ACL, EMNLP, NAACL, COLING, and EACL as top-tier conferences based on low acceptance rates and high citation metrics, but certainly other groupings are also reasonable.
\\[-9pt]


\noindent \textit{Discussion:} Overall, the median citation count is 12. 
75\% of the papers have 34 or fewer citations. 
The average number of citations (45) is markedly higher than the median (12); this is  because of a small number highly cited papers. 

When comparing different types of papers, we notice a large difference between journal papers and the rest.
Even though the number of journal papers in AA (and AA$'$) is very small (about 2.5\%),
these papers have the highest median and average citations (55 and 204, respectively).
Top-tier conferences come next, followed by other conferences. 
The differences between each of these pairs is
statistically significant (Kolmogorov--Smirnov (KS) test, p $<$ .01).\footnote{KS is a non-parametric test that can be applied to compare distributions without needing to make assumptions about the nature of the distributions. Since the citations data is not normally distributed, KS is especially well suited.}
 Interestingly, the workshop papers and the shared task papers have higher medians
and averages than the non-top-tier conferences. These differences are also significant (KS, p $<$ .01).\\[-3pt]

 \begin{figure}[t!]
 \begin{center}
 	\includegraphics[width=\columnwidth]{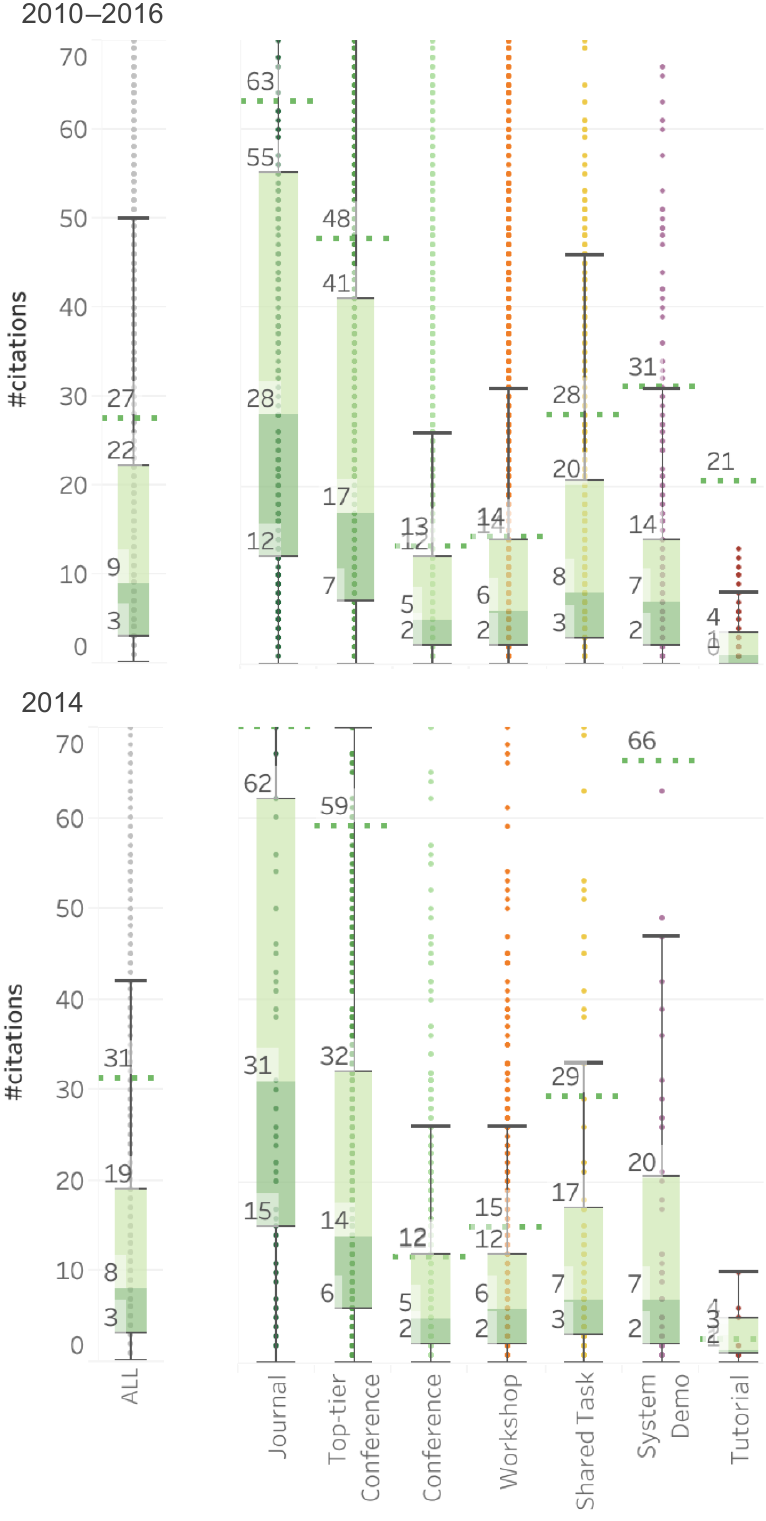}
\vspace*{8mm}
 	\caption{Citation box plots for papers: published 2010--2016 (top) and published in 2014 (bottom).}
 	\label{fig:citn-201016-2014}
 \end{center}
 \end{figure}

\noindent \textit{Q3. How well cited are recent AA$'$ papers: say those published in the last decade (2010--2016)?
How well cited are papers that were all published in the same year, say 2014? Are the citation distributions for individual years very different from those for larger time spans, say 2010--2016?
Also, how well cited are papers 5 years after they are published?}\\[-5pt]

\noindent A. The top of Figure \ref{fig:citn-201016-2014} shows citation box plots for 2010--2016; the bottom shows plots for papers published in 2014.\\[-9pt] 

\noindent \textit{Discussion:} Observe that, in general, these numbers are markedly lower than the those in Figure \ref{fig:citn-overall-ptype}. That is expected as these papers have had less time to accrue citations.

Observe that journal papers again have the highest median and average; however, the gap between journals and top-tier conferences has reduced considerably. The shared task papers have a significantly higher average than workshop and non-top-tier conferences. Examining the data revealed that many of the task description papers
and the competition winning systems' system-description papers received a large number of citations (while the majority of the other system description papers received much lower citations).
Shared tasks have also been particularly popular in the 2010s compared to earlier years.

The plots for 2014 (bottom of Figure \ref{fig:citn-201016-2014}) are similar to that of 2010--2016. (Although, system demo papers published in that year are better cited than the larger set from the 2010--2016 period.)
This plot also gives an idea of citation patterns for papers 5 years after they have been published.\\[-3pt]





 \begin{figure}[t!]
 \begin{center}
 	\includegraphics[width=\columnwidth]{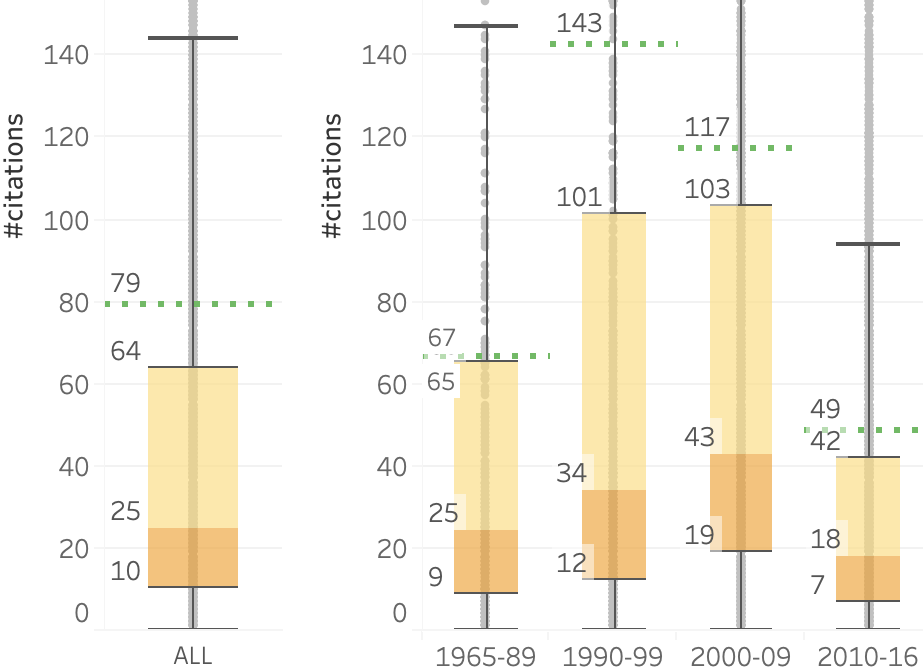}
 	\caption{Citation box plots for journal articles and top-tier conference papers from various time spans.}
 	\label{fig:citn-time}
 \end{center}
 \end{figure}

\noindent \textit{Q4.  If we only consider journal papers and top-tier conferences,
how well cited are papers from various time spans?}\\[-5pt]

\noindent A. Figure \ref{fig:citn-time} shows the numbers for four time spans.\\[-9pt]

\noindent \textit{Discussion:} Observe that the 1990s and the 2000s have markedly higher medians and averages than other time periods. The early 1990s, which have the highest average, were an interesting period for NLP with the emergence of statistical approaches (especially from speech processing) and the use of data from the World Wide Web. 
The 2000--2010 period, which saw an intensification of the statistical data-driven approaches, is notable for
the highest median. 
The high average in the 1990s is likely because of some seminal papers that obtained a very high number of citations.
(Also the 1990's had fewer papers than the 2010s, and thus the average is impacted more by the very high-citation papers.)
The drop off in the average and median for recent papers is largely because they have not had as much time to collect citations.\\[-3pt]





 \begin{figure*}[t!]
 \begin{center}
  \includegraphics[width=2\columnwidth]{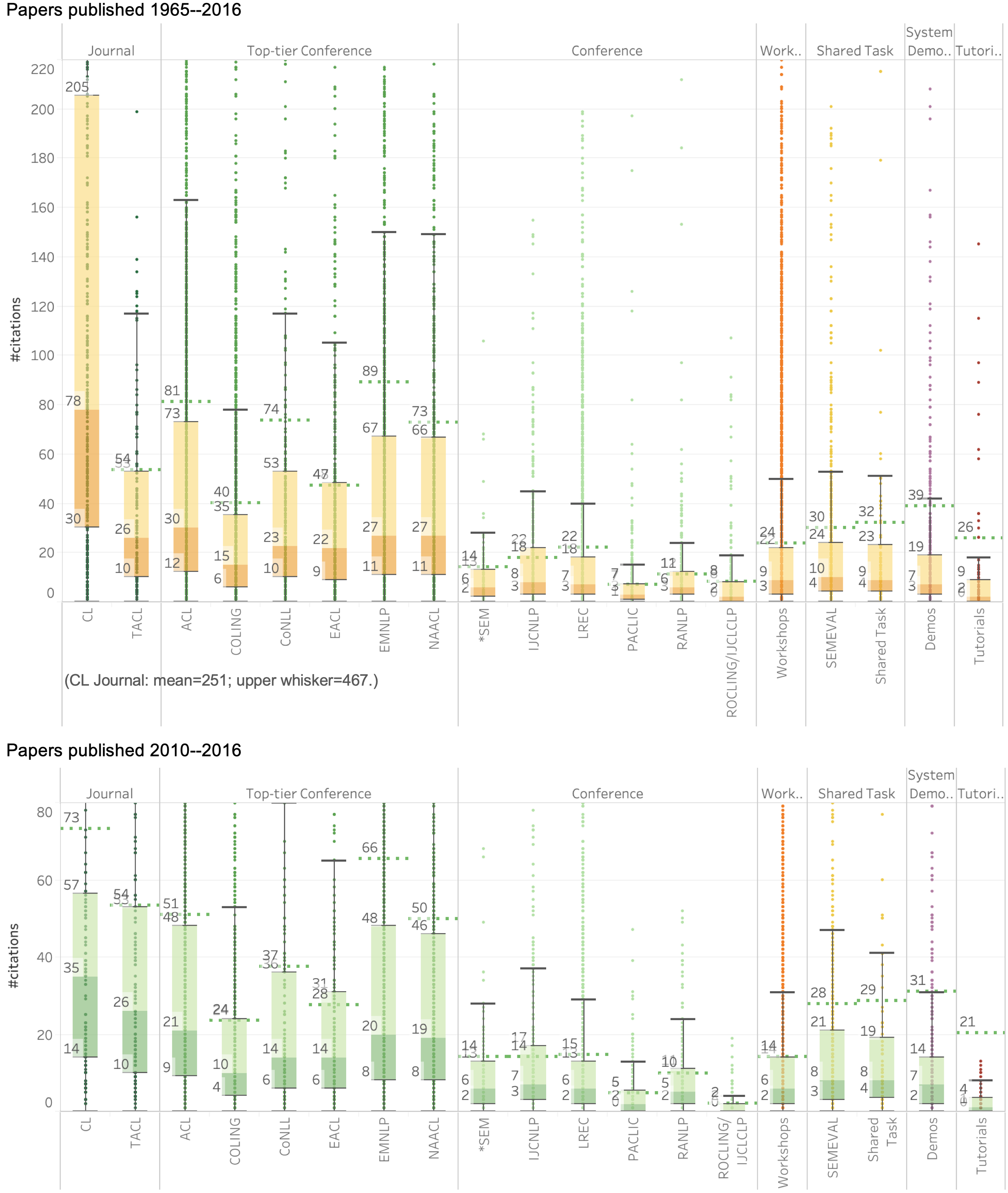}
 	\caption{Citation box plots for papers by venue, type: papers published 1965--2016 (top) and papers published 2010--2016 (bottom).} 
 	\label{fig:citn-venue-ptype}
 \end{center}
 \end{figure*}
 

\noindent Q5. \textit{How well cited are papers from individual NLP venues?}\\[-5pt] 

\noindent A.  Figure \ref{fig:citn-venue-ptype} (top)
shows the citation box plots for 1965--2016 papers from individual venues. 
The plots for workshops,
system, demos, shared tasks, and tutorials are shown as well for ease of comparison.
Figure \ref{fig:citn-venue-ptype} (bottom) 
shows the same box plots for 2010--2016 papers. 

\noindent Discussion: CL Journal has the highest median and average citation numbers. 
ACL comes second, 
closely followed by EMNLP and NAACL.
The gap between CL Journal and ACL is considerably reduced when considering the 2010--2016 papers.
IJCNLP and LREC have the highest numbers among the non-top-tier conferences, but their numbers remain lower than the numbers for SemEval, non-SemEval shared tasks, and workshops.

TACL, a journal, has substantially lower citation numbers
than CL Journal, ACL, EMNLP, and NAACL (Figure \ref{fig:citn-venue-ptype} top). However, it should be noted that TACL only began publishing since 2013. (Also, with a page limit of about ten, TACL papers are arguably more akin to conference papers than journal papers.)
When considering only the 2010--2016 papers, TACL's citation numbers are second only to CL Journal (Figure \ref{fig:citn-venue-ptype} bottom).

When considering 2010--2016 papers, the system demonstration papers, the SemEval shared task papers, and non-SemEval shared task papers have notably high averages (surpassing or equalling those of COLING and EACL); however their median citations are lower. (This is consistent with the trends we saw earlier in Q3.) \\[-3pt]

\begin{figure}[t!]
 \begin{center}
 \includegraphics[width=0.8\columnwidth]{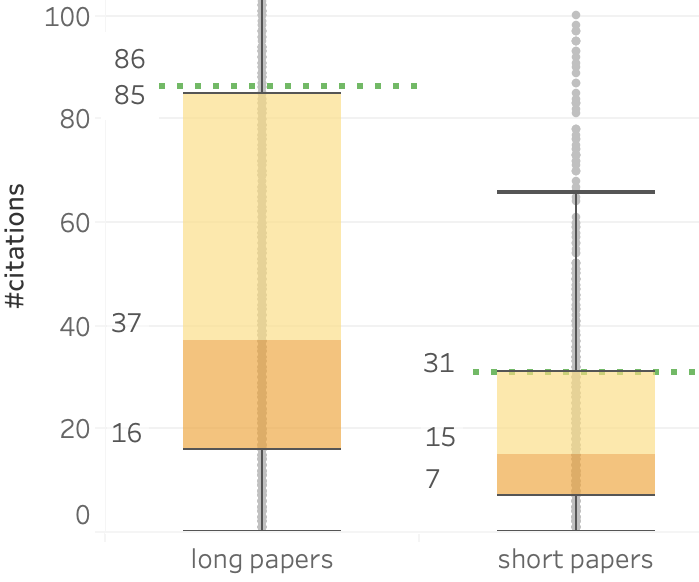}
 \vspace*{6mm}
 	\caption{Citations box plots for long and short ACL papers published between 2003 and 2016.}
 	\label{fig:citn-long-short}
 \end{center}
 \vspace*{3mm}
 \end{figure}

\noindent \textit{Q6. How well cited are long and short ACL main conference papers, respectively?}\\[-5pt]

\noindent A. Short papers were introduced by ACL in 2003. Since then ACL is by far the venue with the highest number of short papers (compared to other venues). So we compare long and short papers published at ACL since 2003 to determine their average citations. 
Figure \ref{fig:citn-long-short} shows the citation box plots for long and short papers published between 2003 and 2016 at ACL. The two distributions are statistically  different (Kolmogorov--Smirnov test, p $<$ .01).\\[-9pt]

\noindent Discussion: In 2003, the idea of short papers was a novelty. It was conceived with the idea that there needs to a be a place for focused contributions that do not require as much space as a long paper. 
The format gained popularity quickly,
and short papers at ACL tend to be incredibly competitive (sometimes having a lower acceptance rate than long papers).
While there have been several influential short papers, it remains unclear how well-cited they are as a category. This analysis sheds some light to that end.
We find that, on average, long papers get almost three times as many citations as short papers; the median for long papers is two-and-half times that of short papers.\\[-5pt]


\noindent \textit{Q7. How do different venues and paper types compare in terms of the volume of papers pertaining to various amounts of citation?}\\[-5pt]

\begin{figure}[t!]
 \begin{center}
 	\includegraphics[width=\columnwidth]{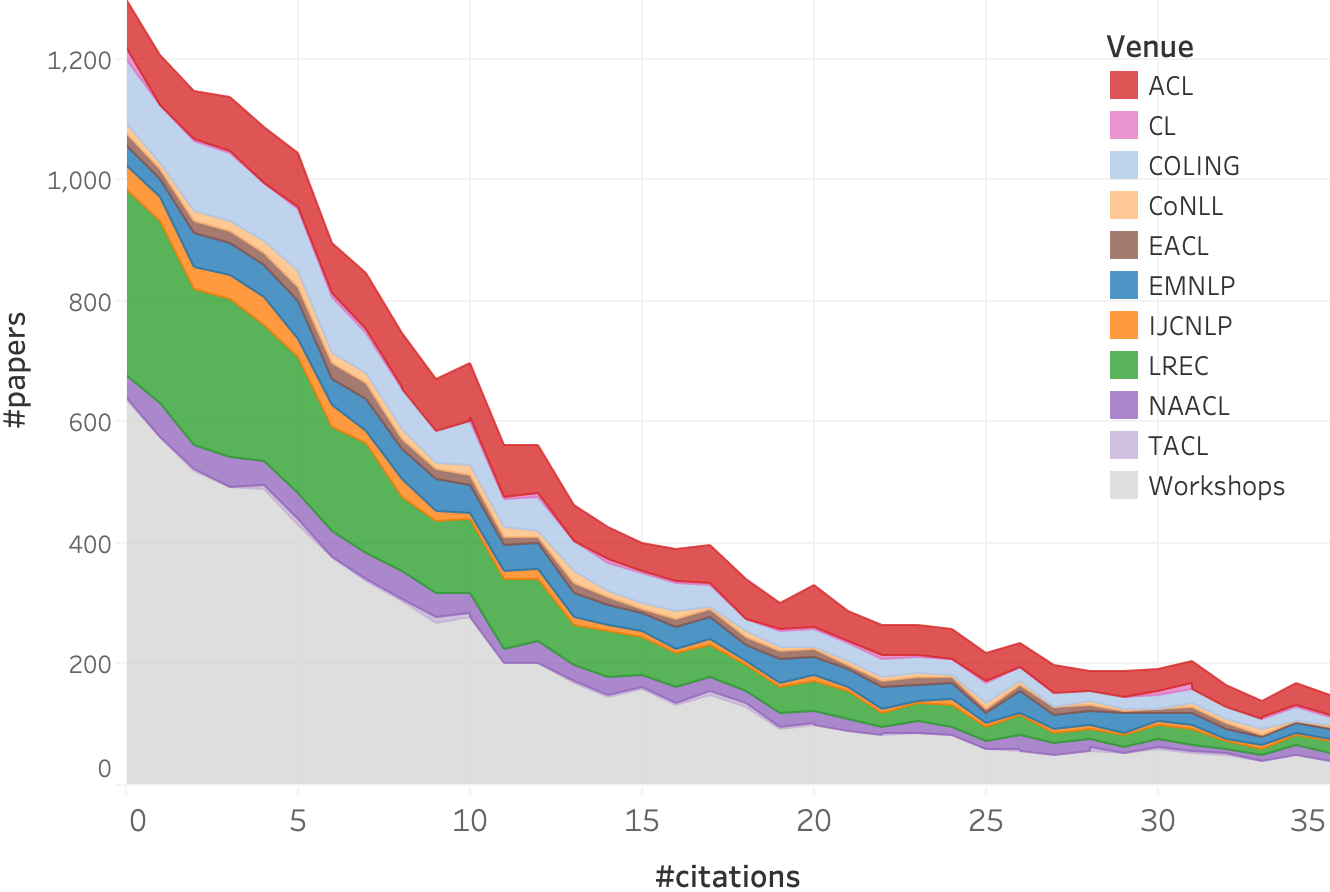}
 \vspace*{1mm}
 	\caption{Stream graph of \#papers by \#citations. The contribution of each venue and paper type is stacked one on top of another.}
 	\label{fig:stream}
 \end{center}
 \end{figure}

\noindent A. Figure \ref{fig:stream} shows a stream graph of \#papers by \#citations. The contributions of each of the venues and paper types are stacked one on top of another (bands of colors). For a given point on the citations axis (say $k$), the width of the stream corresponds to the number of papers with $k$ citations.\\[-9pt]

\noindent \textit{Discussion:} It is not surprising to see that the \#papers by \#citations curve follows a power law distribution. (There are lots of papers with 0 or few citations, but the number drops of exponentially with the number of citations.) Workshop papers (light grey) are the most numerous, followed by LREC (green)---as observable from their wide bands.
The bands for ACL, COLING, EMNLP, and NAACL are easily discernable but the bands for many others, especially CL Journal and TACL are barely discernable indicating low relative volume of their papers.

Observe that the bands for workshops and LREC are markedly wider in the 0 to 10 citations range than in the 11 and more citations range of the \textit{x} axis. In contrast, the widths of the bands for top-tier conferences, such as ACL and EMNLP, remain relatively stable.
Nonetheless, in terms of raw volume, it is worth noting that the workshops and LREC each produce more papers that are cited ten or more times than any other venue. As one considers even higher citations, the top-tier conferences become more dominant.\\[-3pt]

 \noindent \textit{Q8. What percentage of papers 
are cited more than 10 times?\footnote{Google Scholar invented the i-10 index as another measure of author research impact. It stands for the number of papers by an author that received ten or more citations. (Ten here is somewhat arbitrary, but reasonable.)} How many papers are cited 0 times?}\\[-7pt]

\begin{figure}[t!]
 \begin{center}
 	\includegraphics[width=\columnwidth]{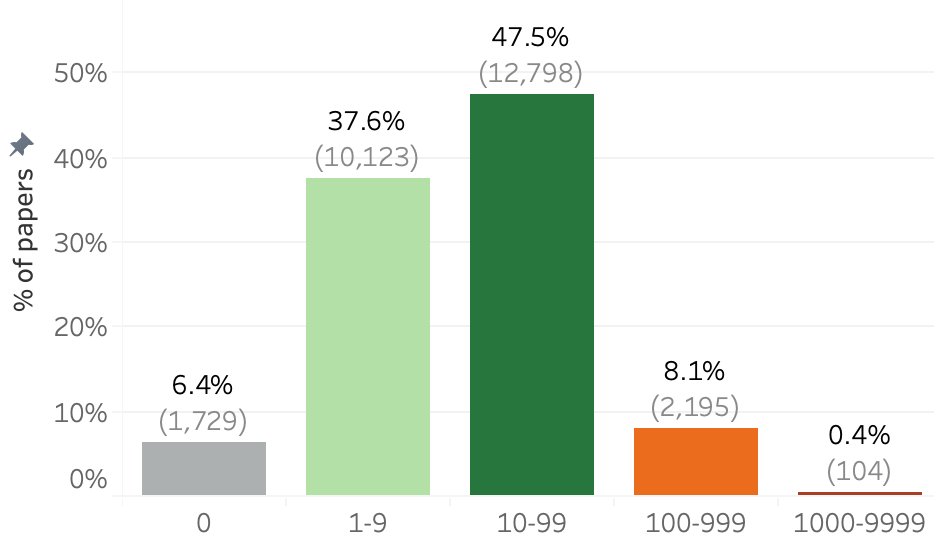}
 	\caption{The percentage of AA$'$ papers in various citation bins. In parenthesis: \#papers.}
 	\label{fig:citnBins-1965-2016}
 \end{center}
 \end{figure}

\noindent A.  
Figure \ref{fig:citnBins-1965-2016} shows the percentage of AA$'$ papers in various citation bins: 0, 1--9, 10--99, and 1000--9999.
(The number of papers is shown in parenthesis.)\\[-9pt]

\noindent Discussion: About 56\% of the papers are cited ten or more times. 6.4\% of the papers are never cited. (Note also that some portion of the 1--9 bin likely includes papers that only received self-citations.)
It would be interesting to compare these numbers with those in other fields such as medical sciences, physics, linguistics, machine learning,
and psychology.\\[-3pt]

 \begin{figure}[t!]
 	\includegraphics[width=0.96\columnwidth]{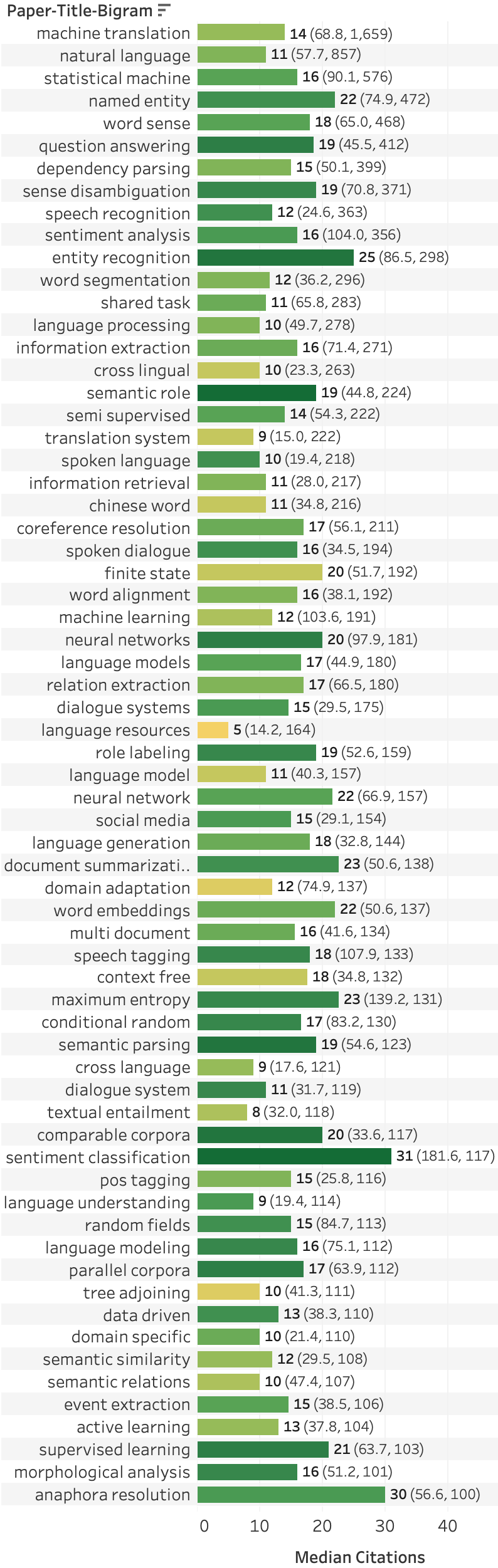}
 	\vspace*{-1mm}
 	\caption{Bar graph of median citations. Title bigrams ordered by number of papers. In parenthesis: average citations, \#papers.} 
 	\label{fig:bigram-citns}
 \vspace*{-9mm}
 \end{figure}

\noindent \textit{Q9. How well cited are areas within NLP?}\\[-5pt]

\noindent A. We used word bigrams in the titles of papers to sample papers from various areas.\footnote{Other approaches such as clustering are also reasonable; however, results with those might not be easily reproducible. We chose the title bigrams approach for its simplicity.} 
The title has a privileged position in a paper. It serves many functions, but most importantly, it conveys what the paper is about.
For example, a paper with the bigram {\it machine translation} in the title is likely about machine translation (MT).
We removed function words from the titles of papers in AA, and extracted all bigrams. 
Figure \ref{fig:bigram-citns} shows, in order of decreasing frequency, the list of 66 bigrams that occurred in more than 100 papers. 
For each bigram, the yellow/green bar shows the median citations of the corresponding papers. The average citations and the number of papers are shown in parenthesis. \\[-14pt] 


\noindent  Discussion: The graph shows, for example, that the bigram \textit{machine translation} occurred in 1,659 AA$'$ papers that have a median citation count of 14, while the average is 68.8. The average is one of the highest among the bigrams, despite the median being more middle of the pack. This suggests 
the presence of heavily cited, outlier, papers. Indeed, the most cited paper in all of AA$'$ is an MT paper with more than 9000 citations \cite{papineni2002bleu}.
Note that not all MT papers have \textit{machine translation} in the title. 
Although non-random, this sample of 1,659 papers is arguably a reasonably representative sample of MT papers.

Third in the list are papers with \textit{statistical machine} in the title---most commonly from the phrase \textit{statistical machine translation}. One expects considerable overlap across these sets of  papers.
However, machine translation likely covers a broader range of research including work done before statistical MT was introduced, as well as work on neural MT and MT evaluation.

The bigrams with the highest median include: \textit{sentiment classification} (31),  \textit{anaphora resolution} (30), and \textit{entity recognition} (25).
The bigrams with the lowest median include: \textit{language resources} (5), \textit{textual entailment} (8), 
\textit{translation system} (9), and \textit{cross language }(9).
The bigrams with the highest average include: \textit{sentiment classification} (181.6), \textit{speech tagging} (107.9), \textit{sentiment analysis} (104.0), and \textit{statistical machine} (90.1).\footnote{Note that
simply composing titles with these high-citation bigrams is not expected to attract a large number of citations.}
One can access the lists of highly cited papers, pertaining to each of the bigrams, through the interactive visualization.

\section{Limitations and Future Work}
We list below some ideas of future work that we did not explore in this paper:\\[-20pt]
\begin{itemize}    
\item Analyze NLP papers that are published outside of the ACL Anthology.\\[-20pt]
\item Measure involvement of the industry in NLP publications over time.\\[-20pt]
\item Measure the impact of research publications in other ways beyond citations. Identify papers that have made substantial contributions in non-standard ways. \\[-20pt]
\end{itemize}
\noindent A list of limitations and ethical considerations associated with this work is available online.\footnote{https://medium.com/@nlpscholar/about-nlp-scholar-62cb3b0f4488}


\section{Conclusions}

We extracted citation information for $\sim$1.1M papers from Google Scholar profiles of researchers who published at least three papers in the ACL Anthology. We used the citation counts of a subset ($\sim$27K papers) to examine patterns of citation across paper types, venues, over time, and across areas of research within NLP.

We showed that only about 56\% of the papers are cited ten or more times.
CL Journal has the most cited papers, but the citation gap between CL journal and top-tier conferences has reduced in recent years.
On average, long papers get almost three times as many citations as short papers.
In case of popular shared tasks, the task-description papers and competition-winning system-description papers often receive a considerable number of citations. 
So much so that the average number of citations for the shared task papers is higher than the average for non-top-tier conferences.
The papers on \textit{sentiment classification},  \textit{anaphora resolution}, and \textit{entity recognition} have the highest median citations.
 Workshop papers and the shared task papers have higher median and average citations than the non-top-tier conferences.

The analyses presented here, and the associated dataset of papers mapped to citations, have a number of uses including,  
understanding how the field is growing and quantifying the impact of different types of papers.
In separate work, we explored the use of the dataset to detect gender disparities in authorship and citations \cite{mohammad2020gender}. The dataset can potentially also be used to compare patterns of citations in NLP with those in other fields. Finally, we note again that citations are not an accurate reflection of the quality or importance of individual pieces of work. A crucial direction of future work is to develop richer ways of capturing scholarly impact.





\section*{Acknowledgments}

This work was possible due to the helpful discussion and encouragement from a number of awesome people including: Dan Jurafsky, Tara Small, Michael Strube, Cyril Goutte, Eric Joanis, Matt Post, Patrick Littell, Torsten Zesch, Ellen Riloff, Iryna Gurevych, Rebecca Knowles, Isar Nejadgholi, and Peter Turney. Also, a big thanks to the ACL Anthology and Google Scholar Teams for creating and maintaining wonderful resources.

\bibliography{ACL2020-Citations_in_NLP}
\bibliographystyle{acl_natbib}

\end{document}